\documentclass[lettersize,journal]{IEEEtran}
\usepackage{amsmath,amsfonts}
\usepackage{algorithmic}
\usepackage{algorithm}
\usepackage{array}

\usepackage{textcomp}
\usepackage{stfloats}
\usepackage{url}
\usepackage{verbatim}
\usepackage{graphicx}
\usepackage{cite}
\usepackage{xcolor}
\usepackage{hyperref} 
\usepackage{subcaption}  

\hypersetup{
    colorlinks=true,   
    linkcolor=blue,    
    citecolor=blue,    
    urlcolor=blue      
}
\usepackage{booktabs}

\newcommand{\rev}[1]{{#1}}
\hypersetup{hidelinks}
\begin{document}
\title{\textcolor{black}{Self-Aligned Resonant Beams for Adaptive Beamforming in 6G IoT}}

\author{Yixuan Guo, Mingliang Xiong, Qingwen Liu, and Wen Fang}
%
\maketitle

\begin{abstract}
\textcolor{black}{The integration of communication, sensing, and wireless power transfer (WPT) is becoming an important direction for \rev{6G Internet of Things (IoT)}. However, conventional beamforming typically depends on channel state information (CSI) acquisition, beam training, and active beam scanning, which can introduce substantial overhead in dynamic or passive-device scenarios. This article reviews the radio-frequency resonant beam system (RF-RBS) as a physical-layer self-alignment mechanism for reducing this dependence. By using retro-directive antenna arrays (RAAs) at the transceivers, RF-RBS can form a cyclic electromagnetic feedback loop in which phase-conjugated waves reinforce a dominant round-trip path. This mechanism enables self-aligned, high-gain beam formation under suitable loop-gain, propagation, and hardware conditions. We discuss the RF-RBS architecture, its operating principles, and its potential roles in WPT, self-aligned communication, and passive positioning. We also examine implementation challenges, including analog hardware integration, multi-user interference, limited coverage, and \rev{line-of-sight (LoS)} dependence, and outline its potential value for latency-sensitive 6G IoT scenarios such as unmanned systems and industrial automation.}
\end{abstract}

\section{Introduction}

\IEEEPARstart{D}{riven} by the demands of 6G, the \rev{Internet of Things (IoT)} is advancing rapidly toward the vision of ``intelligent connectivity of everything.'' Beyond merely enhancing data rates, 6G is expected to deeply integrate communication, positioning, and wireless power transfer, establishing a foundation for ubiquitous intelligent applications.

Regarding integrated communication and sensing, 6G IoT leverages high-frequency bands, massive \rev{multiple-input multiple-output (MIMO)}, and reconfigurable intelligent surfaces (RIS) to achieve centimeter-level high-precision positioning while ensuring high-capacity, low-latency communication~\cite{10380596}. \rev{Radio-frequency (RF)} structures such as RIS can dynamically shape the channel, improving the synergy between positioning and communication~\cite{9874802}.

\textcolor{black}{For the integration of communication and wireless power transfer (WPT), 6G IoT is expected to support simultaneous wireless information and power transfer (SWIPT), which enables concurrent data transmission and wireless energy replenishment for low-power IoT devices, thereby reducing battery replacement and maintenance overhead under favorable link-budget and safety conditions~\cite{10534278}. 
Recent advanced SWIPT designs, such as stacked intelligent metasurface-assisted SWIPT, have further explored electromagnetic-domain beam manipulation for improving information-energy transfer efficiency~\cite{11161741}.}

Empowered by 6G, the integration of communication, positioning, and power transfer in \rev{IoT} is accelerating, greatly enhancing network intelligence and self-organization, and becoming the core foundation for future applications in smart cities, industry, healthcare, and more.

However, in practical integration, challenges in beam control and channel estimation increase system complexity, especially in passive scenarios or low signal quality environments. Traditional beam control methods relying on channel state information (CSI) face two major issues: obtaining accurate CSI is extremely difficult, and in millimeter wave/terahertz bands, the beams are exceedingly narrow, resulting in high overhead for beam measurement and training, which affects beam acquisition and tracking efficiency. In dynamic environments, rapid changes in channel characteristics and frequent user or environmental movement further exacerbate the complexity of beam control. Passive structures like RIS cannot autonomously generate or decode beams and require external control, which may further increase system complexity~\cite{9319211,9964037,9971740}.

Therefore, two core questions are identified for passive scenarios:
\begin{enumerate} 
\item How to improve power transfer efficiency, forming a positive feedback loop that enhances high-quality communication and high-precision positioning. 
\item How to achieve high-precision adaptive beamforming while reducing system complexity. 
\end{enumerate}

\textcolor{black}{Recently, radio-frequency resonant beam system (RF-RBS), enabled by retro-directive antenna arrays (RAAs), has emerged as a promising physical-layer self-alignment mechanism for 6G IoT. 
By exploiting phase conjugation and resonant feedback, RF-RBS can adaptively establish directional beams without explicit CSI acquisition, digital beam training, or active beam scanning. 
Rather than summarizing individual prior studies, this article provides a tutorial-style and perspective-oriented review of RF-RBS, covering its hardware architecture, operating principles, representative roles in WPT, passive positioning, and wireless communication, and its comparison with traditional beamforming. 
We further discuss key opportunities and open challenges.}

\section{RF-RBS: State of the Art}
\subsection{\textcolor{black}{Technique Overview}}

\subsubsection{Wireless Power Transfer}
\textcolor{black}{Current WPT technologies primarily include RF, inductive coupling, and resonant coupling, and have been widely used to power IoT devices. 
However, their practical efficiency is often limited by transmission distance, transmitter-receiver misalignment, environmental loss, and multi-user interference. 
As a result, it remains challenging to provide continuous, stable, and efficient wireless charging for IoT terminals. 
Moreover, dynamic multi-device scenarios further increase the need for proper power-node deployment, accurate beam control, and efficient energy scheduling.}

\subsubsection{Indoor Positioning}
\textcolor{black}{Mainstream indoor positioning technologies, such as \rev{radio-frequency identification (RFID), ultra-wideband (UWB),} and Wi-Fi, can achieve high accuracy in specific environments. 
However, multipath, blockage, and interference make it difficult to maintain stable centimeter-level accuracy in complex indoor scenarios, especially for passive positioning without terminal cooperation. 
High-precision positioning also depends on high-quality communication links and complex signal processing, which increase energy consumption and limit real-time responsiveness.}

\subsubsection{Wireless Communication}
\textcolor{black}{In millimeter-wave, terahertz, and cooperative beamforming systems, beam control and sidelobe suppression remain key challenges. 
Accurate beam pointing requires real-time adaptation to user mobility, multipath propagation, and hardware imperfections. 
Meanwhile, severe sidelobes may cause energy leakage, inter-user interference, \rev{signal-to-noise ratio (SNR)} degradation, capacity loss, and potential security risks. 
Therefore, stable communication in dynamic IoT scenarios requires both accurate main-lobe alignment and effective interference control.}

\subsubsection{Interplay and Coupling}
\textcolor{black}{WPT, indoor positioning, and wireless communication are tightly coupled in IoT systems. 
Stable communication links improve positioning accuracy, while accurate positioning helps optimize beam pointing and resource allocation. 
However, both high-precision positioning and high-quality communication increase terminal energy consumption, making reliable wireless power supply essential. 
Therefore, improving WPT efficiency is fundamental to sustaining communication and positioning performance in complex IoT environments.}

\begin{figure*}[!t]
  \centering
  \includegraphics[width=0.7\textwidth]{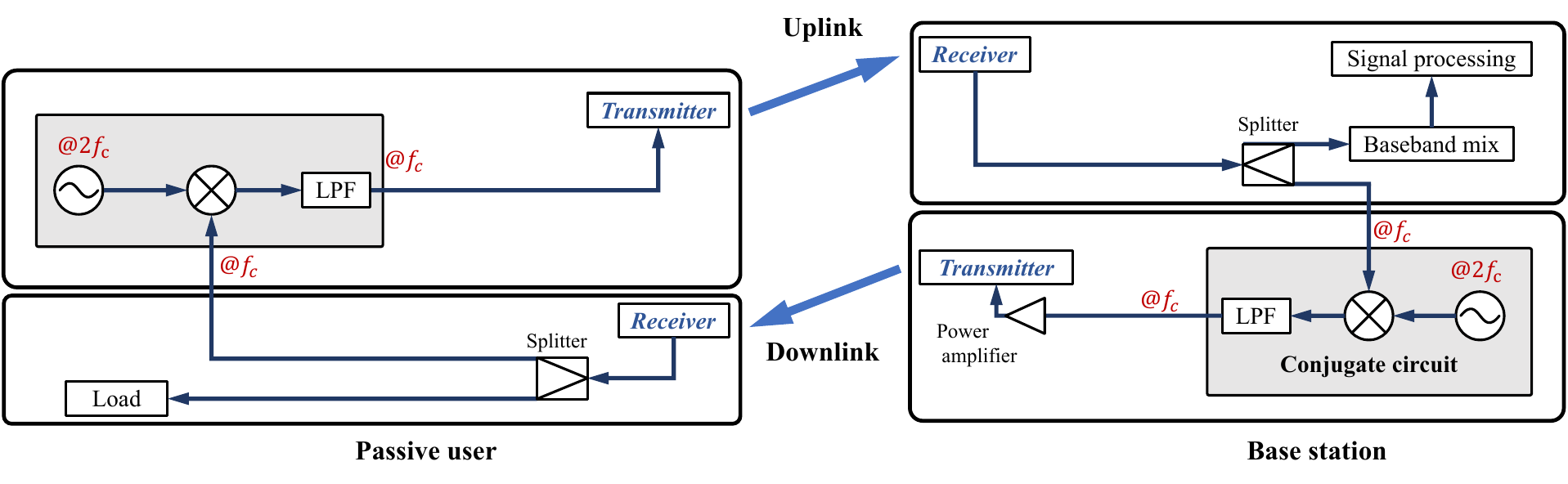}  
  \caption{Hardware architecture and transceiver circuit design of the RF-RBS.}
  \label{rbs}
\end{figure*}

\subsection{Structure and Principles}
The RF-RBS is a novel RF wireless transmission architecture based on the \textcolor{black}{RAAs}. Its core mechanism lies in deploying RAAs with phase conjugation capability at both the base station (BS) and the user equipment (UE) ends, thereby constructing a self-sustaining round-trip cyclic electromagnetic wave path at the physical layer.
When the BS and UE are spatially aligned, the incident wave is automatically phase conjugated by the UE's RAA and reflected back to the BS along the original path. This conjugated echo coherently superimposes with the BS's local excitation signal in the feeding network, forming a positive feedback loop that drives the system into a resonant locking state in a specific spatial direction. In this state, electromagnetic energy is efficiently cyclically accumulated between the BS and UE, spontaneously generating a highly directional, {self-aligning} resonant beam without the need for CSI estimation or active beam scanning algorithms.

It is worth noting that while the steady-state operation is self-sustaining, the initial activation of the passive tag typically requires a low-power wake-up signal or a rectified voltage from the ambient RF environment to cold-start the conjugation circuit before the high-gain resonant loop is fully established.

Regarding the system architecture illustrated in Fig.~\ref{rbs}, both the BS and UE integrate phase conjugation circuits based on heterodyne mixing: the received signal is mixed with a local oscillator signal with twice the carrier frequency (\(f_{\text{LO}}=2f_{\text{c}}\)), and then the phase-conjugated wave (\(\phi \to -\phi\)) is output after \textcolor{black}{band}-pass filtering. 
A key distinction lies in the BS architecture, which incorporates a power amplifier to compensate for the losses during the entire system operation and ensure the continuous operation of the system.

\subsection{\textcolor{black}{Operating Conditions and Resonant Formation}}

\textcolor{black}{RF-RBS should be understood as a condition-dependent physical-layer mechanism rather than a universal replacement for all beamforming systems. 
A stable resonant beam can be established only when the UE-side conjugation circuit is activated by stored energy, a low-power wake-up signal, or rectified RF energy and the round-trip resonant loop provides sufficient gain to compensate for the accumulated propagation and RF-chain losses over repeated iterations. 
These losses may come from free-space attenuation, power splitting, phase-conjugation circuits, imperfect amplification, and array mismatch. 
If the per-iteration loop gain is insufficient, the circulating signal gradually decays and the resonant beam cannot be sustained.}

\begin{figure}[!t]
  \centering
  \includegraphics[width=0.7\linewidth]{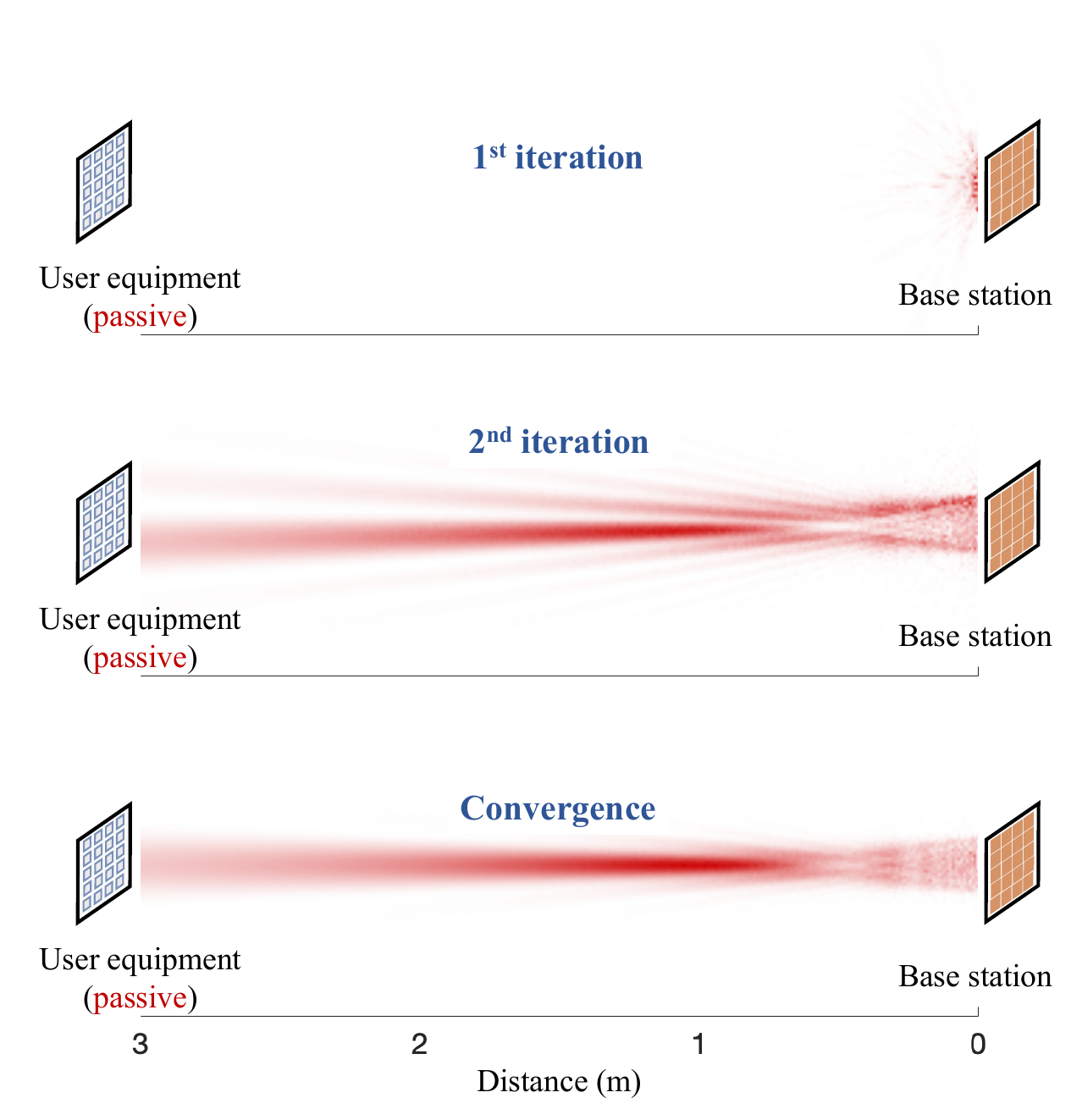}  
  \caption{\textcolor{black}{Power convergence process and spatial power distribution of the resonant beam.}}
  \label{fig:iteration}
\end{figure}

\textcolor{black}{The formation of a stable beam also requires a dominant reciprocal \rev{line-of-sight (LoS)} or strong specular path. 
In a multipath environment, different propagation components may act as candidate resonant modes. 
The component with higher round-trip efficiency and better phase consistency is gradually reinforced through repeated phase-conjugated feedback, whereas lossy or phase-mismatched components decay over iterations. 
This mode-selection process explains why RF-RBS can concentrate energy along the dominant path, as illustrated by the convergence behavior and spatial power distribution in Fig.~\ref{fig:iteration}.}

\textcolor{black}{Nevertheless, severe multipath, blockage, or closely spaced reflective users may create competing resonant modes and degrade beam stability. 
Hardware and geometric nonidealities, such as oscillator phase noise, UE rotation, and UE tilt, may also reduce phase-conjugation accuracy and weaken resonant locking. 
Therefore, RF-RBS is most effective in short-range or infrastructure-assisted IoT deployments where a stable dominant round-trip path can be maintained.}

\subsection{\textcolor{black}{Wireless Power Transfer}}

\textcolor{black}{Although RAAs have been explored for WPT by automatically redirecting energy toward a beacon source~\cite{9921326}, conventional RAA-based WPT mainly relies on one-sided retro-direction rather than a self-sustained bidirectional resonant loop.}
The RF-RBS was applied to the domain of WPT in \cite{10855572}, specifically targeting key IoT bottlenecks such as low transfer efficiency and complex beam control. 
\textcolor{black}{In contrast, RF-RBS uses the closed-loop resonant formation mechanism discussed above to realize self-aligned energy concentration between the BS and UE.}
The proposed system allows users to shunt and collect a portion of energy after receiving signals, and feed back a small amount of power to the BS; the BS transmits again after power amplification. 
\textcolor{black}{Through repeated phase-conjugated feedback, the dominant energy-transfer path is reinforced, leading to a steady-state resonant field distribution.}
This not only \textcolor{black}{reduces off-target energy leakage} and phase distortion in non-target directions but also significantly improves the spatial power focusing degree.

\textcolor{black}{From a WPT perspective, this closed-loop energy focusing mechanism provides a potential route for cable-free power delivery and simultaneous information transfer in short-range IoT scenarios.}
Simulation verification demonstrates that in the 30 GHz frequency band, the proposed scheme can stably output watt-level \rev{direct-current (DC)} power and support a downlink spectral efficiency of 21 bps/Hz; its self-alignment capability has \textcolor{black}{good robustness to moderate} three-dimensional offsets, making it suitable for cable-free energy supply requirements in dynamic IoT scenarios. 
\textcolor{black}{Nevertheless, the practical power level and transfer efficiency remain dependent on link distance, RF-chain loss, rectifier efficiency, safety constraints, and hardware implementation.}

\subsection{Indoor Positioning}
\textcolor{black}{RF-RBS has also been extended to passive indoor positioning scenarios~\cite{10636970}.} By utilizing the self-established resonant beam between a single BS and a passive UE as the detection wave, there is no need for the target to actively transmit signals or for the system to perform complex beam scanning and channel estimation. The BS can achieve high-precision direction of arrival (DoA) estimation via a \rev{multiple signal classification (MUSIC)} algorithm. This is also mainly attributed to the inherent power focusing and self-alignment robustness of the resonant beam, which significantly improves the echo SNR. Building on this, a triangulated \textcolor{black}{RF-RBS} was proposed to advance the passive positioning scheme from two-dimensional (2D) direction finding to three-dimensional (3D) positioning \cite{11005386}. 

\textcolor{black}{It should be noted that such millimeter-level positioning accuracy is obtained under favorable LoS-dominant and array-calibrated conditions, rather than being guaranteed in all indoor environments. 
The MUSIC-based DoA estimation relies on a high-SNR dominant resonant path and an accurate array manifold. 
Owing to the resonant feedback and mode-selection mechanism, RF-RBS can partially suppress weak multipath components and small random phase perturbations, since components with lower round-trip efficiency or inconsistent phase evolution tend to decay over iterations. 
However, strong reflectors, complete \rev{non-line-of-sight (NLoS)} blockage, or large inter-element phase errors may still create competing resonant modes and bias the estimated DoA. 
In addition, array gain/phase mismatch, element-position errors, mutual coupling, excessive \rev{phase-locked loop (PLL)} jitter, and synchronization errors may distort the steering vector and degrade the subspace separation required by MUSIC. 
Therefore, practical indoor deployment requires careful array calibration, phase-noise control, and LoS or strong-specular-path maintenance.}

\subsection{Wireless Communication}

\textcolor{black}{Although retro-directive backscatter systems have been explored for communication and localization by redirecting incident signals toward their sources~\cite{kludze2024frequency}, RF-RBS communication is different because it must carry bidirectional information while maintaining a repeated closed-loop resonance.
When a single-frequency RF-RBS is used for communication, the uplink and downlink share the same carrier during round-trip propagation, leading to severe echo interference. }
To address this issue, a dual-frequency separation design was proposed in \cite{11220199}. 
By assigning different carrier frequencies to the uplink and downlink, frequency division duplexing (FDD) is realized while retaining the phase conjugation capability of the RAAs, which effectively avoids the severe echo interference problem in traditional single-frequency RBS. 
Its core lies in replacing the local oscillator frequency of \(2f_{\text{c}}\) in the ``conjugation circuit'' shown in Fig.~\ref{rbs} with a local oscillator frequency equal to the sum of the uplink and downlink carrier frequencies (\(f_{\text{tx}}+f_{\text{rx}}\)) at the input of the ``conjugation circuit''. 
Meanwhile, by separating the transmitting and receiving arrays at the BS and MT ends, the element spacing is ensured to satisfy \(\lambda_{\text{tx}}/d_{\text{tx}} = \lambda_{\text{rx}}/d_{\text{rx}}\), so that the dual-frequency signals can still be accurately reflected along the original path and maintain strong coupling resonance.

Further expanding this architecture, a SWIPT system was developed to \textcolor{black}{support} a spectral efficiency of 4.8 bps/Hz alongside watt-level power transfer within indoor ranges \cite{10660556}. 

\subsection{\textcolor{black}{Performance}}

\begin{table*}[htbp]
\renewcommand{\arraystretch}{1.08}
\centering
\small
\caption{\rev{Key tradeoffs: RF-RBS and conventional pilot-based beamforming.}}
\label{tab:rbs_vs_traditional}
\begin{tabular}{>{\raggedright\arraybackslash}p{3cm}>{\raggedright\arraybackslash}p{6.6cm}>{\raggedright\arraybackslash}p{6.6cm}}
\toprule
\rev{\textbf{Characteristic}} & \textbf{RF-RBS} & \rev{\textbf{Conventional beamforming}} \\
\midrule
\rev{Beam alignment} & \rev{Phase-conjugated feedback} & \rev{Pilot/codebook-based control} \\
\rev{Alignment overhead} & \rev{No explicit CSI or beam scanning} & \rev{Training and/or feedback} \\
\rev{Beam response} & \rev{Microseconds (numerical example)} & \rev{Protocol- and hardware-dependent} \\
\rev{Processing burden} & \rev{Less digital alignment; added analog circuitry} & \rev{Architecture-dependent} \\
Energy efficiency & \rev{Link- and RF-loss-dependent} & \rev{Array- and precoder-dependent} \\
\rev{Motion and blockage} & \rev{Self-alignment; blockage/tilt-sensitive} & \rev{Update-based tracking; blockage-sensitive} \\
\rev{Coverage} & \rev{Round-trip-loss-limited} & \rev{Wider-area capability} \\
\rev{Hardware} & \rev{Conjugation and loop control} & \rev{Mature implementations} \\
\bottomrule
\end{tabular}
\end{table*}

\begin{figure}[t]
  \centering
  \begin{subfigure}{0.8\linewidth}
    \centering
    \includegraphics[width=\linewidth]{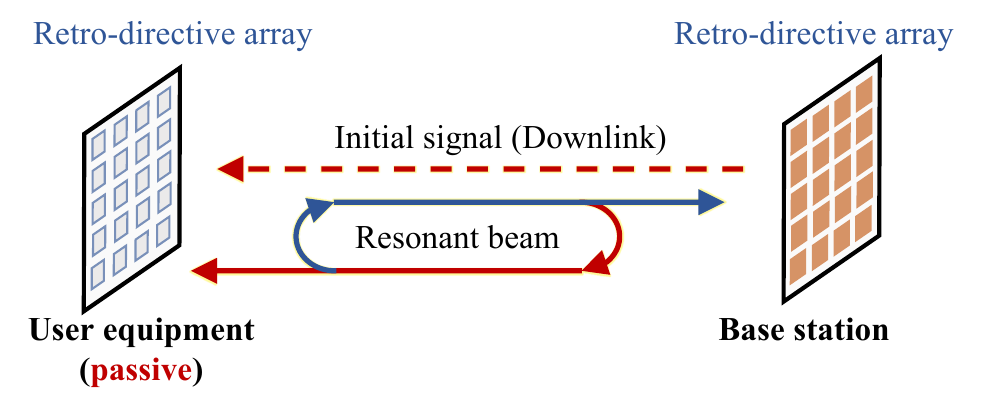}
    \caption{\textcolor{black}{RAAs} adaptive beamforming.}  
    \label{subfig:rda}          
  \end{subfigure}
  \hfill  
  \begin{subfigure}{0.8\linewidth}
    \centering
    \includegraphics[width=\linewidth]{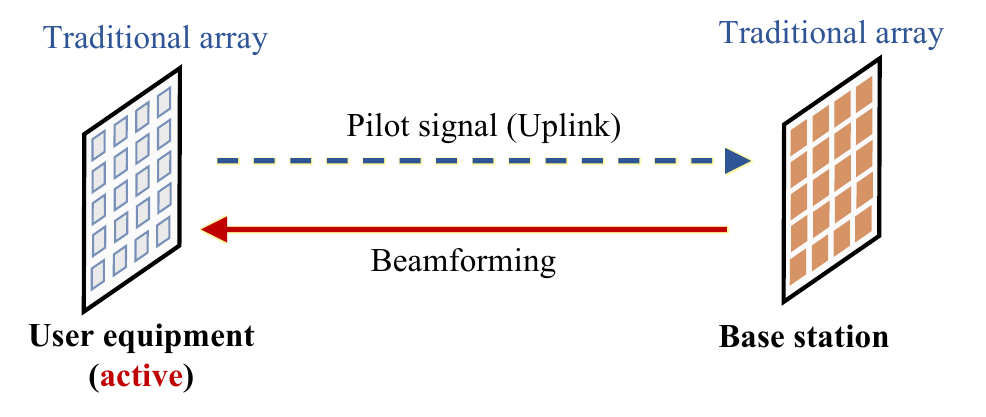}
    \caption{Traditional array beamforming.}
    \label{subfig:ta}
  \end{subfigure}
  \caption{Schematic comparison of beamforming principles: RF-RBS vs. Traditional active phased array}
  \label{fig:bf}  
\end{figure}

\textcolor{black}{Table~\ref{tab:rbs_vs_traditional} compares RF-RBS with traditional pilot-based beamforming. 
\rev{Pilot-based} beamforming \rev{may require} CSI estimation, beam training, \rev{or precoder updates}, whereas RF-RBS uses physical phase conjugation for adaptive beam alignment, as illustrated in Fig.~\ref{fig:bf}. 
\rev{RF-RBS enables self-alignment without explicit CSI estimation or active beam scanning, reducing beam-management overhead for dynamic short-range IoT links.}}

\begin{figure}[!t]
    \centering
    \includegraphics[width=0.8\linewidth]{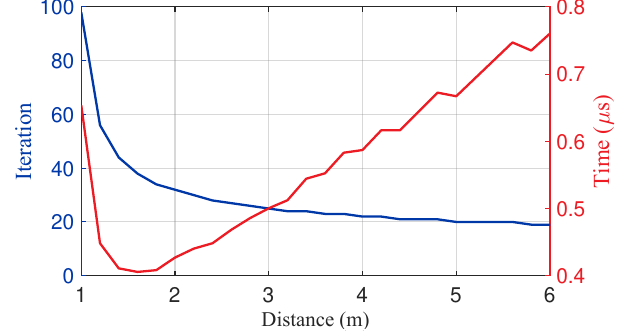}
    \caption{\textcolor{black}{Iteration number and convergence time versus distance.}}
    \label{fig:iter_time}
\end{figure}

\textcolor{black}{\rev{Conventional beamforming latency depends on the protocol, hardware, CSI feedback, and beam-training overhead}~\cite{8458146}. 
RF-RBS instead forms the beam through physical round-trip iterations, so its response time is determined by the required iteration number and the per-iteration RF/analog loop delay~\cite{10660556}. 
As shown in Fig.~\ref{fig:iter_time}, RF-RBS \rev{shows microsecond-level beam convergence} over the considered transmission distances \rev{in this numerical example}. 
Meanwhile, the required iteration number decreases as the distance increases. 
This is mainly because short-distance transmission suffers from stronger near-field effects, including larger wavefront curvature and element-wise phase variations, and therefore requires more iterations for phase self-adaptation and resonant-mode selection. 
At longer distances, the wavefront becomes closer to planar, so fewer iterations are needed for beam stabilization. 
\rev{This numerical result supports rapid beam formation under the considered loop-gain, hardware, and propagation conditions.}}

\textcolor{black}{Moreover, RF-RBS has limited coverage and higher hardware cost: the former comes from the need to sustain repeated round-trip resonance under propagation and RF-chain losses, while the latter comes from additional phase-conjugation and loop-control hardware. \rev{RF-RBS thus enables low-overhead beam alignment for dynamic short-range IoT.}}

\section{Applications in Next-Generation IoT}
\begin{figure*}[!t]
  \centering
  \includegraphics[width=0.65\textwidth]{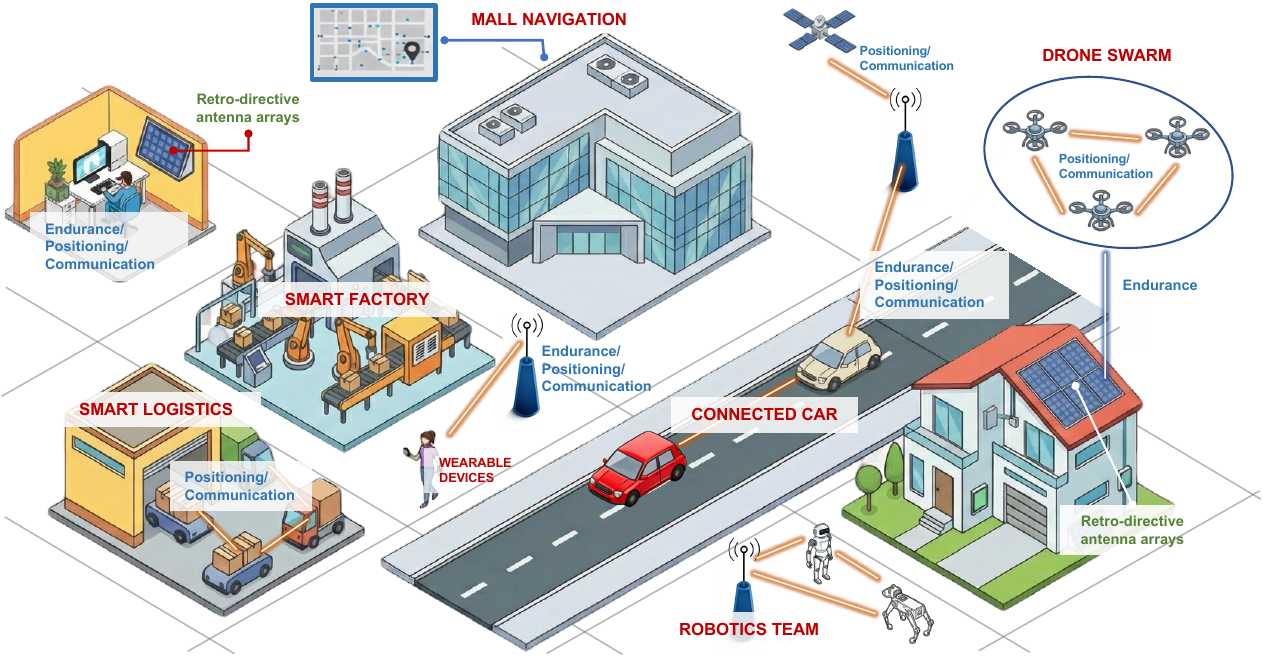}  
  \caption{Typical emerging application scenarios enabled by RF-RBS in 6G smart city.}
  \label{app}
\end{figure*}

\textcolor{black}{The RF-RBS technology, leveraging its physical-layer self-alignment mechanism, can potentially support three coupled functions in 6G IoT: wireless power transfer, indoor positioning, and wireless communication. 
This section discusses representative IoT scenarios where self-aligned beam formation may reduce beam-control overhead and support integrated communication, positioning, and energy supply under suitable deployment conditions. Typical emerging application scenarios are illustrated in Fig.~\ref{app}.}

\subsection{Unmanned Systems}
\textcolor{black}{\rev{Artificial intelligence (AI)-enabled} unmanned systems impose stringent requirements on link stability, positioning accuracy, and device endurance. 
For smart vehicles, drone swarms, and robotic teams, high mobility can easily cause beam misalignment and link interruption. 
By exploiting physical phase conjugation, RF-RBS can support fast self-aligned beam tracking without complex CSI estimation or active beam scanning, thereby improving link stability in dynamic short-range scenarios.}

\textcolor{black}{Accurate autonomous decision-making also requires reliable positioning support. 
By establishing a self-sustaining resonant beam link with the BS, RF-RBS can provide high-precision passive positioning under favorable link and calibration conditions.
Meanwhile, its WPT capability offers a potential cable-free energy replenishment method for energy-constrained devices such as drones, although the practical power level depends on link distance, safety constraints, and hardware efficiency.}

\subsection{Industrial IoT}
\textcolor{black}{Industrial IoT is moving toward autonomous and flexible operation, but metal obstacles, moving equipment, and dense devices can introduce multipath interference and unstable links. 
Conventional passive sensing technologies such as RFID and Wi-Fi may suffer from limited positioning accuracy in complex factory environments, while wired power supply restricts the mobility of industrial robots and other devices.}

\textcolor{black}{RF-RBS can support self-aligned cable-free communication and power transfer by locking onto target devices through resonant feedback. 
Its passive positioning capability can also assist item tracking and device identification on production lines. 
Therefore, RF-RBS is more suitable for industrial scenarios that require reliable connectivity, flexible deployment, and reduced maintenance, while its performance still depends on LoS availability, array calibration, and hardware integration.}

\subsection{Smart Living}
\textcolor{black}{Smart living scenarios, including smart homes, mall navigation, and wearable devices, require reliable indoor connectivity, positioning, and device endurance. 
RF-RBS can provide cable-free energy replenishment for low-power IoT devices and use the resonant beam as a probing signal for indoor positioning where \rev{the Global Positioning System (GPS)} is unavailable, such as malls, underground areas, or complex indoor environments.}

\textcolor{black}{For human-centered applications such as wearable devices, safety-compliant operation is a prerequisite. 
The RF-RBS operation depends on the balance between loop gain and loss; when the effective resonant link cannot be maintained, the circulating power naturally decays. 
This property may help reduce unnecessary radiation outside the intended operating region, but practical deployment still requires compliance with RF exposure regulations and careful power-control design.}

\textcolor{black}{Overall, these application scenarios suggest that RF-RBS is particularly suitable for IoT devices with moderate hardware capability rather than arbitrary ultra-low-cost tags. 
In short-range infrastructure-supported deployments, its benefits in self-aligned communication, wireless energy replenishment, and positioning may partially offset the added hardware cost. 
Practical adoption still depends on the tradeoff among application value, device cost, hardware complexity, and maintenance requirements.}

\begin{figure*}[!t]
  \centering
  \includegraphics[width=0.7\textwidth]{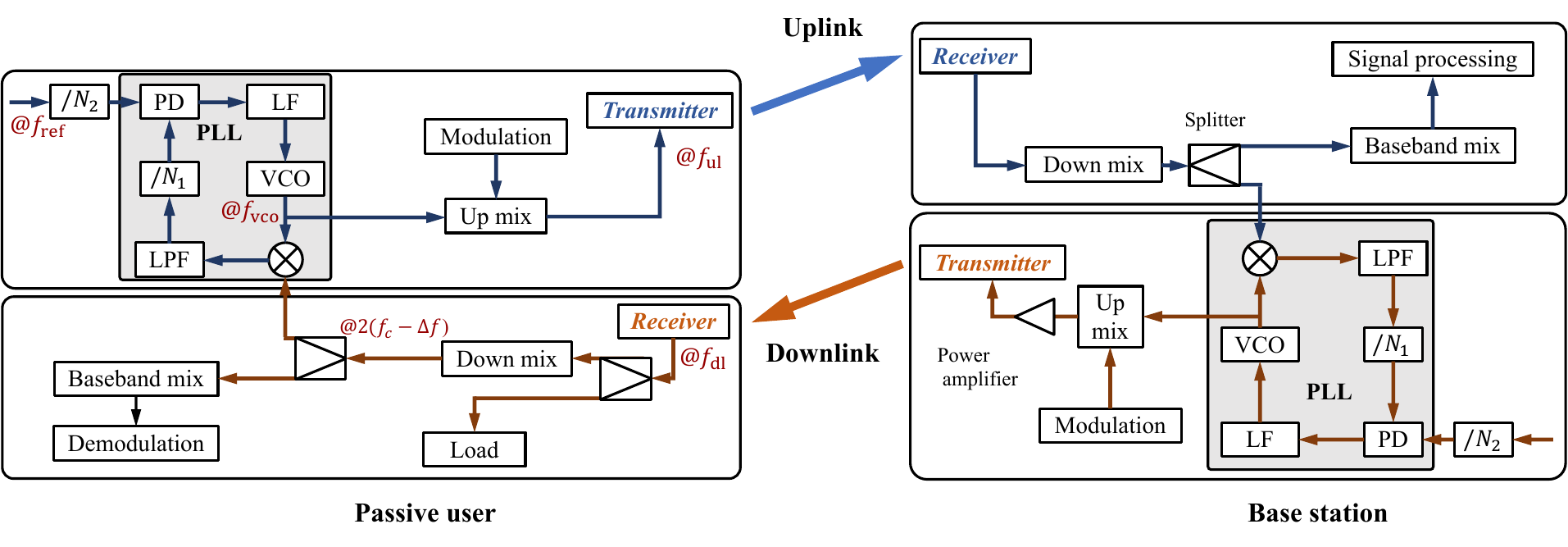}  
  \caption{Schematic diagram of the PLL-based FDD architecture.}
  \label{rbcom}
\end{figure*}

\section{Opportunities and Challenges}
\subsection{Opportunities}
\subsubsection{Native ISAC Architecture}
As 6G evolves towards the integration of ``Communication, Sensing, and Energy'', RF-RBS provides a native physical-layer solution. Unlike traditional solutions that simply superimpose communication and sensing functions, RF-RBS relies on a unified resonant beam mechanism to simultaneously achieve high-precision positioning and efficient data transmission, enabling deep functional integration. \textcolor{black}{Furthermore, by reducing the need for explicit CSI estimation and active beam scanning, RF-RBS may lower part of the digital beam-control burden. However, this benefit is obtained by shifting complexity to the analog RF front end, so its suitability for massive passive IoT depends on hardware integration, cost, and power consumption}.

\subsubsection{\textcolor{black}{Wireless Energy Replenishment for Low-Power IoT}}
\textcolor{black}{The WPT capability of RF-RBS offers a potential path to alleviating the battery-life challenge for low-power IoT devices. 
Through the design of a phase-conjugated positive feedback loop, this technology has been shown to support watt-level DC power output under favorable indoor link conditions. 
This power supply capability provides a potential route toward battery-assisted or maintenance-light IoT sensing, particularly in scenarios with stringent endurance and maintenance requirements, such as industrial monitoring and smart homes.}

\subsubsection{Intrinsic Physical Layer Security}
\textcolor{black}{RF-RBS provides a promising physical-layer security opportunity by forming a spatially selective resonant beam between the BS and UE. 
For passive eavesdroppers or low-power jammers located outside the dominant resonant path, the repeated phase-conjugated feedback concentrates signal power around the intended direction, thereby reducing off-axis information leakage and weakening off-path interference. 
This spatial selectivity makes RF-RBS attractive for security-sensitive IoT scenarios such as industrial control, smart logistics, and private indoor networks.}
\textcolor{black}{This opportunity is most relevant when a stable LoS or strong specular resonant path is maintained. 
If an adversary is close to the main path, located near strong scatterers, or able to inject a spoofed phase-conjugated response, the resonant loop may be disturbed. 
In multi-user deployments, energy competition and resonant-mode aliasing among closely spaced reflective devices should also be considered. 
Therefore, RF-RBS can serve as a physical-layer security enhancement, while authentication, encryption, and interference management remain necessary for complete security protection.}

\subsubsection{Robustness in High-Mobility Scenarios}
Leveraging its rapid response mechanism based on circuit physics, RF-RBS demonstrates \textcolor{black}{potential} advantages in \textcolor{black}{highly} dynamic scenarios. 
\rev{The beam-convergence time of RF-RBS} \textcolor{black}{is expected to reach the microsecond level under suitable loop-gain and hardware conditions}. 
This \textcolor{black}{may help mitigate} Doppler effect and rapid link switching issues in high-speed mobile environments. 
This characteristic gives it \textcolor{black}{promising} potential for application in vehicle-to-everything (V2X) and drone swarm collaboration, which require \textcolor{black}{low-latency} communication and stability, offering a new solution for maintaining links in dynamic environments.

\subsection{Challenges}

\subsubsection{Analog Hardware Integration}
\textcolor{black}{The RF-RBS shifts system complexity to the analog domain, making compact and low-power RF front-end integration a key challenge. 
Practical deployment requires integrating phase-conjugation circuits, mixers, filters, \rev{PLLs}, and frequency-conversion modules into power-constrained IoT devices. 
Although FDD can mitigate same-frequency echo interference by separating the uplink and downlink bands, it requires careful filtering, frequency synthesis, and \rev{RF-to-intermediate-frequency (IF)} isolation.}
A promising solution is the PLL-based frequency division architecture shown in Fig.~\ref{rbcom}. 
In this scheme, the PLL synthesizes the local oscillator signal locked to the reference phase. 
By flexibly configuring the frequency divider ratios ($N_1$ and $N_2$), the system can independently generate the required uplink ($f_\text{ul}$) and downlink ($f_\text{dl}$) carrier frequencies. 
This approach enables simultaneous frequency switching and phase conjugation, thereby enhancing system flexibility.

\textcolor{black}{The corresponding data flow follows a two-branch process. 
After receiving the RF signal, the front-end down-converts it to an intermediate-frequency signal. 
One branch is converted to baseband for data demodulation, while the other enters the PLL path for carrier phase recovery and frequency synthesis. 
The outgoing data are then modulated onto the regenerated carrier and retransmitted through the phase-conjugation path, preserving the resonant round-trip phase relationship.}

\subsubsection{Multi-User Interference Management}
Interference in a multi-user environment is a challenging problem for traditional solutions. In the RF-RBS system, ensuring that multiple users can simultaneously excite resonant beams without mutually competing for energy or causing frequency aliasing still necessitates support from refined multiple access protocols.  A primary technical hurdle lies in distinguishing different reflective sources and \textcolor{black}{establishing} independent resonant links when multiple users are closely spaced.

Lightweight scaling solutions that coordinate bandwidth, hardware, and algorithms need to be explored, such as hybrid multiple access, sparse array beam reuse, or AI-based dynamic resource allocation strategies, to achieve efficient and scalable deployment.
To address the ``energy competition'' among multiple users, future designs may incorporate unique spectral signatures or code-domain modulation within the retro-directive loop, allowing the BS to physically distinguish and selectively amplify resonant beams for specific target users.

\subsubsection{Range and Coverage Extension}
The operation of the RF-RBS system relies on the loop gain being greater than the path loss. Resonance ceases once the distance exceeds a specific threshold, a challenge that is exacerbated by the extremely high path loss in millimeter-wave or terahertz bands. Expanding the effective resonant distance without violating safety radiation standards is a key challenge for extending its application range. Simultaneously, although the link can be self-established when the UE moves in RF-RBS, this highly depends on the relative attitude of the array. When the UE rotates or tilts relative to the BS, its effective receiving aperture shrinks, and the spatial phase distribution of the reflected signal becomes distorted, which can also lead to severe degradation of phase conjugate accuracy and significant attenuation of resonance intensity. Future mitigation should be achieved through attitude sensing and control optimization.

\subsubsection{High Dependency on LoS}
The establishment of the resonant beam relies on the physical cyclic path between the transmitter and the receiver. If the path is completely blocked, the positive feedback loop breaks, and communication and power transfer immediately cease. 
While this is considered an ``inherently secure'' characteristic, it also means poor service continuity in complex occlusion environments. Promising solutions may need to combine technologies such as RIS for relay assistance to address \rev{NLoS} transmission issues. However, how to use RIS to guide the BS transmission signal toward a passive target and establish a resonant link with it may require further consideration.

\section{CONCLUSION}
The RF-RBS leverages physical-layer phase conjugation to achieve a native integration of communication, sensing, and power transfer. Characterized by its self-alignment and self-sustaining capabilities, \textcolor{black}{the system can support self-aligned beam formation while reducing the need for complex digital baseband processing.}
This mechanism significantly mitigates the latency and overhead associated with CSI acquisition, \textcolor{black}{making it a promising candidate for selected dynamic 6G IoT scenarios}.

However, widespread adoption relies on addressing inherent implementation hurdles, particularly regarding analog hardware integration, multi-user interference, and LoS dependencies. Future research focused on overcoming these physical constraints through engineering optimization or hybrid architectures remains a critical step toward realizing the commercial potential of this low-complexity paradigm.
\bibliographystyle{IEEEtran}

\bibliography{Mybib}

@ARTICLE{9874802,
  author={Chen, Rui and Liu, Mengjie and Hui, Yilong and Cheng, Nan and Li, Jiandong},
  journal={IEEE Internet of Things Journal}, 
  title={Reconfigurable Intelligent Surfaces for 6{G} {IoT} Wireless Positioning: A Contemporary Survey}, 
  year={2022},
  volume={9},
  number={23},
  pages={23570-23582},
  doi={10.1109/JIOT.2022.3203890}}

@ARTICLE{10380596,
  author={Mu, Xidong and Xu, Jiaqi and Liu, Yuanwei and Hanzo, Lajos},
  journal={IEEE Vehicular Technology Magazine}, 
  title={Reconfigurable Intelligent Surface-Aided Near-Field Communications for {6G}: Opportunities and Challenges}, 
  year={2024},
  volume={19},
  number={1},
  pages={65-74},
  doi={10.1109/MVT.2023.3345608}}

@ARTICLE{10534278,
  author={Psomas, Constantinos and Ntougias, Konstantinos and Shanin, Nikita and Xu, Dongfang and Mayer, Kenneth and Tran, Nguyen Minh and Cottatellucci, Laura and Choi, Kae Won and Kim, Dong In and Schober, Robert and Krikidis, Ioannis},
  journal={Proceedings of the IEEE}, 
  title={Wireless Information and Energy Transfer in the Era of {6G} Communications}, 
  year={2024},
  volume={112},
  number={7},
  pages={764-804},
  doi={10.1109/JPROC.2024.3395178}}

@ARTICLE{9964037,
  author={Meng, Xi and Zhang, Nan and Jian, Mengnan and Kadoch, Michel and Yang, Dacheng},
  journal={IEEE Internet of Things Journal}, 
  title={Channel Modeling and Estimation for Reconfigurable-Intelligent-Surface-Based {6G SAGIN IoT}}, 
  year={2023},
  volume={10},
  number={11},
  pages={9273-9282},
  doi={10.1109/JIOT.2022.3224847}}

@ARTICLE{9971740,
  author={Fang, Xinran and Feng, Wei and Chen, Yunfei and Ge, Ning and Zhang, Yan},
  journal={IEEE Internet of Things Journal}, 
  title={Joint Communication and Sensing Toward 6{G}: Models and Potential of Using {MIMO}}, 
  year={2023},
  volume={10},
  number={5},
  pages={4093-4116},
  doi={10.1109/JIOT.2022.3227215}}

@ARTICLE{9319211,
  author={López, Onel L. A. and Alves, Hirley and Souza, Richard Demo and Montejo-Sánchez, Samuel and Fernández, Evelio Martin Garcia and Latva-Aho, Matti},
  journal={IEEE Internet of Things Journal}, 
  title={Massive Wireless Energy Transfer: Enabling Sustainable IoT Toward {6G} Era}, 
  year={2021},
  volume={8},
  number={11},
  pages={8816-8835},
  doi={10.1109/JIOT.2021.3050612}}

@ARTICLE{10855572,
  author={Jiang, Qingwei and Liu, Mingqing and Xu, Mengyuan and Fang, Wen and Xiong, Mingliang and Liu, Qingwen and Zhou, Shengli},
  journal={IEEE Internet of Things Journal}, 
  title={Single-Frequency Self-Alignment {RF} Resonant Beam for Information and Power Transfer}, 
  year={2025},
  volume={12},
  number={11},
  pages={16622-16636},
  doi={10.1109/JIOT.2025.3535147}}

@ARTICLE{10636970,
  author={Guo, Yixuan and Jiang, Qingwei and Xu, Mengyuan and Fang, Wen and Liu, Qingwen and Yan, Gang and Yang, Qunhui and Lu, Hai},
  journal={IEEE Transactions on Wireless Communications}, 
  title={Resonant Beam Enabled {DoA} Estimation in Passive Positioning System}, 
  year={2024},
  volume={23},
  number={11},
  pages={16290-16300},
  doi={10.1109/TWC.2024.3439703}}

@ARTICLE{11005386,
  author={Guo, Yixuan and Xiong, Mingliang and Fang, Wen and Jiang, Qingwei and Xu, Mengyuan and Liu, Qingwen and Yan, Gang},
  journal={IEEE Internet of Things Journal}, 
  title={Resonant Beam Enabled Passive {3-D} Positioning}, 
  year={2025},
  volume={12},
  number={15},
  pages={30242-30253},
  doi={10.1109/JIOT.2025.3570320}}

@ARTICLE{11220199,
  author={Xia, Shuaifan and Liu, Qingwen and Jiang, Qingwei and Fang, Wen and Liu, Mingqing},
  journal={IEEE Transactions on Wireless Communications}, 
  title={Frequency Division Duplexing Resonant Beam Communication}, 
  year={\textcolor{blue}{2026}},
  volume={\textcolor{blue}{25}},
  number={},
  pages={\textcolor{blue}{6145-6159}},
  doi={10.1109/TWC.2025.3623890}}

@ARTICLE{10660556,
  author={Xia, Shuaifan and Jiang, Qingwei and Fang, Wen and Liu, Qingwen and Zhou, Shengli and Liu, Mingqing and Xiong, Mingliang},
  journal={IEEE Internet of Things Journal}, 
  title={Millimeter-Wave Resonant Beam {SWIPT}}, 
  year={2024},
  volume={11},
  number={24},
  pages={40464-40477},
  doi={10.1109/JIOT.2024.3452121}}

@ARTICLE{8458146,
  author={Giordani, Marco and Polese, Michele and Roy, Arnab and Castor, Douglas and Zorzi, Michele},
  journal={IEEE Communications Surveys \& Tutorials}, 
  title={A Tutorial on Beam Management for {3GPP NR} at {mmWave} Frequencies}, 
  year={2019},
  volume={21},
  number={1},
  pages={173-196},
  doi={10.1109/COMST.2018.2869411}}

@INPROCEEDINGS{11161741,
  author={Amiri, Mojtaba and Javadi, Sepideh and Zarini, Hosein and Mili, Mohammad Robat and An, Jiancheng and Sookhak, Mehdi and Krikidis, Ioannis},
  booktitle={ICC 2025 - IEEE International Conference on Communications}, 
  title={Stacked Intelligent Metasurface for Simultaneous Wireless Information and Power Transfer}, 
  year={2025},
  volume={},
  number={},
  pages={4583-4588},
  doi={10.1109/ICC52391.2025.11161741}}

@ARTICLE{9921326,
  author={Kang, Yuxin and Lin, Xian Qi and Li, Yong and Wang, Bao},
  journal={IEEE Antennas and Wireless Propagation Letters}, 
  title={Dual-Frequency Retrodirective Antenna Array With Wide Dynamic Range for Wireless Power Transfer}, 
  year={2023},
  volume={22},
  number={2},
  pages={427-431},
  doi={10.1109/LAWP.2022.3215207}}

@article{kludze2024frequency,
  title={A frequency-agile retrodirective tag for large-scale sub-terahertz data backscattering},
  author={Kludze, Atsutse and Kono, Junichiro and Mittleman, Daniel M and Ghasempour, Yasaman},
  journal={Nature communications},
  volume={15},
  number={1},
  pages={8756},
  year={2024},
  publisher={Nature Publishing Group UK London}
}

\section*{Biographies}
\vspace{-30pt}

\begin{IEEEbiographynophoto}{Yixuan Guo}
\textcolor{black}{(guoyixuan9676@163.com) received his B.E. degree in software engineering from Shanxi University, Taiyuan, China, in 2018, his M.E. degree in software engineering from Northwest Normal University, Lanzhou, China, in 2022, and his Ph.D. degree in intelligent science and technology from Tongji University, Shanghai, China, in 2026
He is currently a \rev{Associate Professor} with the College of Electronic Information Engineering, Taiyuan University of Technology, Taiyuan, China. His research interests include wireless communications, indoor positioning, wireless power transfer.}
\end{IEEEbiographynophoto}
\vspace{-30pt}

\begin{IEEEbiographynophoto}{Mingliang Xiong}
[M](mlx@tongji.edu.cn) received the B.E. degree in communications engineering from the Nanjing University of Posts and Telecommunications, Nanjing, China, in 2017, and the Ph.D. degree from the College of Electronics and Information Engineering, Tongji University, Shanghai, China, in 2022. He served as a research fellow of Hangzhou Institute of Extremely-Weak Magnetic Field Major National Science and Technology Infrastructure, Hangzhou, China, in 2022-2024. He is currently an Assistant Professor of Tongji University. 
 His research interests include optical wireless communications, wireless power transfer, and the Internet of Things. 
\end{IEEEbiographynophoto}
\vspace{-30pt}

\begin{IEEEbiographynophoto}{Qingwen Liu}
 [SM] (qliu@tongji.edu.cn) received the B.S. degree in electrical engineering and information science from the University of Science and Technology of China, Hefei, in 2001, and the M.S. and Ph.D. degrees from the Department of Electrical and Computer Engineering, University of Minnesota, Minneapolis, in 2003 and 2006, respectively. He is currently a Professor with the College of Computer Science and Technology, Tongji University, Shanghai, China. 
His research interests lie in the areas of wireless communications and signal processing.
\end{IEEEbiographynophoto}
\vspace{-30pt}

\begin{IEEEbiographynophoto}{Wen Fang}
\textcolor{black}{[M](wen.fang@tongji.edu.cn) received the B.E. degree in computer science and technology from the Shandong University of Science and Technology, Qingdao, China, in 2017, and the Ph.D. degree from the College of Electronics and Information Engineering, Tongji University, Shanghai, China, in 2022. From 2022 to 2024, she was a Postdoctoral Research Fellow with the School of Electronic Information and Electrical Engineering, Shanghai jiao Tong University, Shanghai, China. She is currently an Assistant Professor at the College of Electronics and Information Engineering, Tongji University, Shanghai, China. 
 Her research interests include optical wireless communications, the development of remote wireless charging technology, and the Internet of Things.}
\end{IEEEbiographynophoto}

\vfill 

\end{document}